\documentclass[a4paper,11pt]{article}
\usepackage{pos}
\usepackage[export]{adjustbox}

\title{The search for light dark matter with DAMIC-M}

\author*[a]{Radomir Smida}

\affiliation[a]{Enrico Fermi Institute and Kavli Institute for Cosmological Physics, University of Chicago,\\
Chicago, USA}

\note{on behalf of the DAMIC-M Collaboration}

\emailAdd{smida@kicp.uchicago.edu}

\abstract{
The DAMIC-M (DArk Matter In CCDs at Modane) experiment will use skipper CCDs to search for low mass (sub-GeV) dark matter underground at the Laboratoire Souterrain de Modane (LSM) in France. With about 1\,kg of silicon target mass and sub-electron energy resolution, the detector will surpass the exposure and threshold (eV-scale) of previous experiments. Thus, DAMIC-M will have world-leading sensitivity to a variety of “hidden sector” candidates. In this talk, we will report on science results from a prototype detector, test performance of CCD modules, and the status of the detector construction at LSM.
}

\FullConference{42nd International Conference on High Energy Physics (ICHEP2024)\\
18-24 July 2024\\
Prague, Czech Republic\\}

\begin{document}
\maketitle

\section{Introduction}

The DAMIC-M (DArk Matter In CCDs at Modane) experiment has been specifically designed to detect both nuclear and electronic recoils from light (sub-GeV) DM scattering. Due to the smaller mass of such DM particles, the kinematics favor lower energy recoils that require detectors to be sensitive to sub-keV energy deposits, or only few eV for the absorption of dark photons. DAMIC-M will operate deep underground at LSM and contain up to 208 charge-coupled devices (CCDs) for a total sensitive mass of Si of $\sim$700\,g. DM-induced ionization events in the CCD bulk can be detected with sub-electron resolution through nondestructive, repeated pixel readout\,\cite{Janesick1990,Chandler1990,Sensei2017}. The ability to detect single electrons with low dark current sensors in a low background environment ($<$1\,dru = 1\,event/keV/kg/day), will allow DAMIC-M to push the energy threshold down to only a few eV.

In this proceeding, we describe the performance and calibration of DAMIC-M skipper CCDs, the technical details and results of the DAMIC-M detector and its prototype at LSM, the Low Background Chamber (LBC)\,\cite{DAMIC-M:2024ooa}, and future plans.

\section{Skipper charge-coupled devices (CCDs)}

Skipper CCDs developed by Lawrence Berkeley National Laboratory (LBNL), USA\,\cite{Holland2023} were manufactured by Teledyne DALSA Semiconductor, Canada on 150\,mm diameter wafers. High-resistivity ($>$10\,k$\Omega$-cm) and ultra-pure Si crystal was pulled by Topsil GlobalWafers A/S, Denmark. The CCD thickness is 670\,$\mu$m and the device can be fully depleted with the substrate voltage V$_{\rm{sub}}\geq$40\,V. The pixels on the front surface are 15\,$\mu$m\,$\times$\,15\,$\mu$m. Wirebonding and packaging are done in a clean room at the University of Washington.

The CCD has a three-phase polysilicon gate structure with a buried p-type channel, where charge carriers collected from fully depleted high-resistivity ($\>$10\,k$\Omega$-cm) n-type silicon bulk are clocked toward a readout amplifier. Flex cables wire bonded to the CCD provide the required voltage biases and clocks.

DAMIC-M CCDs have floating-gate (called skipper) amplifiers allowing multiple non-destructive charge measurements (NDCMs) by moving pixel charge between the summing well and floating gate. The readout noise then decreases as $\sigma_{1}/\sqrt{N_{\rm{skip}}}$, where $\sigma_1$ is the noise measured for the first readout and $N_{\rm{skip}}$ is the number of NDCMs or skips. Typically, $N_{\rm{skip}}\approx500$ is required to achieve good single-electron resolution ($\sigma<0.2$\,e$^{-}$). Even if the multi-skip readout takes long time, this does not constitute an issue if the skipper CCD takes data in a low background environment with low dark current. In addition, the readout time can shorten by summing charge from more pixels in either vertical or horizontal direction, or both (i.e. binning).

The CCD has excellent spatial reconstruction due to its pixelization and the depth can be reconstructed thanks to charge diffusion happening between the interaction point and the buried channel. The diffusion in the CCD is calibrated by atmospheric muons on the surface. The type of a particle interacting in the CCD can be identified, if the particle leaves a multi-pixel track in the CCD, see\,Ref.\,\cite{DAMIC:2015ipv}. The analysis of spatial and time coincidences is used to measure $^{32}$Si and $^{210}$Pb decays in the CCD\,\cite{DAMIC:2020wkw}. A comprehensive radioactive background model\,\cite{DAMIC:2021crr} of the previous CCD-based experiment, DAMIC at SNOLAB, was constructed and it is being used for DAMIC-M with appropriate activities of contaminants.

Several calibration measurements were done with skipper CCDs during last few years. Their goal is to study the CCD response to gammas and neutrons at very low energies ($<$100\,eV), where we expect the DM signal. These measurements include precision measurement of Compton scattering in Si down to 23\,eV with a $^{241}$Am source\,\cite{Compton2022} and measurement of the nuclear recoil ionization efficiency in Si with low-energy neutrons ($<$24\,keV) from a $^{124}$Sb-$^9$Be photoneutron source (ongoing analysis). Lastly, the possibility to distinguish nuclear recoil signals from electronic recoil backgrounds in CCDs has been studied\,\cite{McGuire:2023cbg}.

\section{DAMIC-M detector}

The design goals of the DAMIC-M experiment are as follows: collect an exposure of $\sim$1\,kg-year, read out pixel charge with single electron resolution, achieve low background and low dark current ($<$0.5\,e$^{-}$/mm$^2$/day). All this will allow DAMIC-M to detect both nuclear and electronic recoils and explore wide range of DM models, i.e. DM masses starting from m$_{\chi}\leq$1.2\,eV (hidden-sector mediators) to light WIMPs (m$_{\chi}<10$\,GeV). 

The detector will be located at LSM in France with 1700\,m of rock overburden or 4800\,m.w.e. reducing the muon flux to only 5\,$\mu$/m$^2$/day. A clean room of class ISO\,5 was built for the DAMIC-M detector and other room will be used for testing CCDs and assembly. Since early 2022, the LBC takes data at LSM. DAMIC-M installation is expected in early 2025.

The DAMIC-M CCD has 6k$\times$1.5k pixels and weighs 3.3\,grams. Four of these devices are epoxied and wire-bonded on a Si pitch adapter, which has traces bringing clock and bias voltages from a CCD flex cable. CCD modules are going to be assembled in an array and connected to cold copper inside the cryostat. The modules are surrounded by an infrared shielding to reduce dark current while being operated at 130\,K. The cryostat is surrounded by 20\,cm of external lead and 30\,cm of high-density polyethylene (HDPE). Custom low-noise electronics for CCDs has been developed and shows promising performance in the LBC. Electronics, slow control (vacuum, temperature, etc.) and data acquisition systems will be operated remotely while being monitored with an automatic data quality system.

To reduce the radioactive background, DAMIC-M has employed various mitigation steps. First, the CCDs will accumulate only 100\,days\footnote{The cosmogenic activation will be $\sim20\times$ lower than in other experiments.} of the total surface equivalent exposure which reduces the amount of activated isotopes in the Si bulk like $^{3}$H, $^{7}$Be and $^{22}$Na\,\cite{Saldanha:2020ubf}. This has been achieved by strict avoidance of transport by air, using a container with 16-ton shielding during transports, underground storage, and having expedite production. To minimize surface contamination, a strict control of the exposure to air with Radon and dust has been adopted. Ultra-clean CCD flex cable with $^{238}$U ($^{232}$Th) content 100 (20) times lower than in standard ones\,\cite{Arnquist:2023gtq} will be glued on a Si pitch adapter at least 1\,cm away from the CCDs. Copper electro-formed (EF) and machined underground at Sanford Underground Research Facility (SURF), USA and Canfranc Underground Laboratory (LSC), Spain, will be used for parts holding the CCD modules, IR shield surrounding the CCD array and the cryostat can. The inner most 5\,cm of the lead shielding will be made from ancient lead. The remaining parts of the cryostat will be made from ultra-pure oxygen-free high thermal conductivity (OFHC) copper -- used already in the LBC. All copper and ancient lead pieces will be chemically cleaned to remove surface contamination like machining oil, grease, dust, etc.. The design has been validated with Geant4 simulations and the expected total background is less than 1\,dru.

\section{LBC detector and results}

The LBC is a DAMIC-M prototype detector at LSM and was commissioned in early 2022. Its goals are as follows: characterize DAMIC-M components, primarily CCDs, in a low background environment; test other subsystems (electronics, SC, DAQ, data transfer, etc.); gain working experience at LSM; and produce science results. Because the detector design is described in\,\cite{DAMIC-M:2024ooa}, we give only a brief overview here. The original setup used two 6k$\times$4k pixel skipper CCDs housed in a box made from OFHC copper. The CCD box is surrounded by 2\,cm of ancient lead and additional 4-10\,cm of low background lead in a cryostat made from OFHC copper. The outer shielding consists of 15\,cm lead and 20\,cm HDPE. A custom data acquisition system based on a commercial controller was used to run CCDs. CCDs in this setup, measured the event rate of (1.1$\pm$0.1)\,ev/g/day above 1\,keV or (12.5$\pm$2.8)\,dru between 1 and 6 keV\,\cite{DAMIC-M:2024ooa}.

The setup described above was used to collect data on DM-electron scattering\,\cite{DAMIC-M:2023gxo,DAMIC-M:2023hgj}. Two data sets were taken during the LBC commissioning phase over three months in 2022 when the resolution was 0.2\,e$^{-}$ with N$_{\rm{skip}}=650$ and dark current $\sim$20\,e$^{-}$/mm$^2$/day. The steps of our data cleaning procedure consisted of the minimum bias criteria to identify high energy clusters and reject those with $>$7\,e$^{-}$; the mask around clusters for charge transfer inefficiency which removes only $6\times10^{-5}$ of the pixels; the rejection of hot pixels and columns; and exclusion of regions with charge traps in the serial register. The integrated exposure for the DM search after implementing all the cuts is 85.2\,g-days\,\cite{DAMIC-M:2023gxo}.

To place an upper limit on the DM signal, a joint binned-likelihood fit is performed on the pixel charge distributions of two CCD amplifiers. The DM signal is computed with QEdark\,\cite{QEdark} for the DM density profile in the galactic halo as recommended in\,\cite{Baxter:2021pqo} after adding diffusion, pixelization and dark current. The constraints on ultra-light and heavy mediators were the best limits at the time of publication\,\cite{DAMIC-M:2023gxo}.

To improve the LBC limit in the 1\,e$^-$ bin, i.e. DM masses below $\sim$3\,MeV, we took advantage of the stability of measured signal in a subset of the LBC data used in the previous work. In this subset, the images were taken consecutively every 10\,min. during 63 days. Due to the scattering in Earth’s bulk, the DM flux and velocity distribution get modified and we expect to see a daily modulation of the DM signal. Because no signal modulation was found for periods of 1-48\,h, we enhanced our previous limit by two orders of magnitude in the DM mass range 0.53-2.7\,MeV\,\cite{DAMIC-M:2023hgj}.

The LBC continues to run as a low background setup for skipper CCDs and several upgrades have been implemented since the middle 2022. First, two prototype CCD modules with devices from the DAMIC-M pre-production run were installed. These new CCDs are less activated than the 6k$\times$4k ones in the previous setup and we can measure impurities in the same Si crystal as is used for the final DAMIC-M CCDs. Second, both lids of the CCD box were exchanged for EF copper from LSC\,\cite{Borjabad:2018wda}. After these two changes, the background rate decreased to (0.9$\pm$0.03)\,ev/g/day above 1\,keV or (6.7$\pm$1.1)\,dru between 1 and 6 keV\,\cite{DAMIC-M:2024ooa}. In addition, new electronics developed for DAMIC-M was installed in early 2024 reducing the readout noise and also the dark current thanks to clock shaping and other modifications.
Further progress has been made on the data analysis, where we focus on the search of $^{32}$Si and $^{210}$Pb coincidences and low energy clusters\,\cite{DAMIC:2021crr}, among others.

\begin{figure*}[!t]
    \centering
    \includegraphics[width=0.45\textwidth, valign=m]{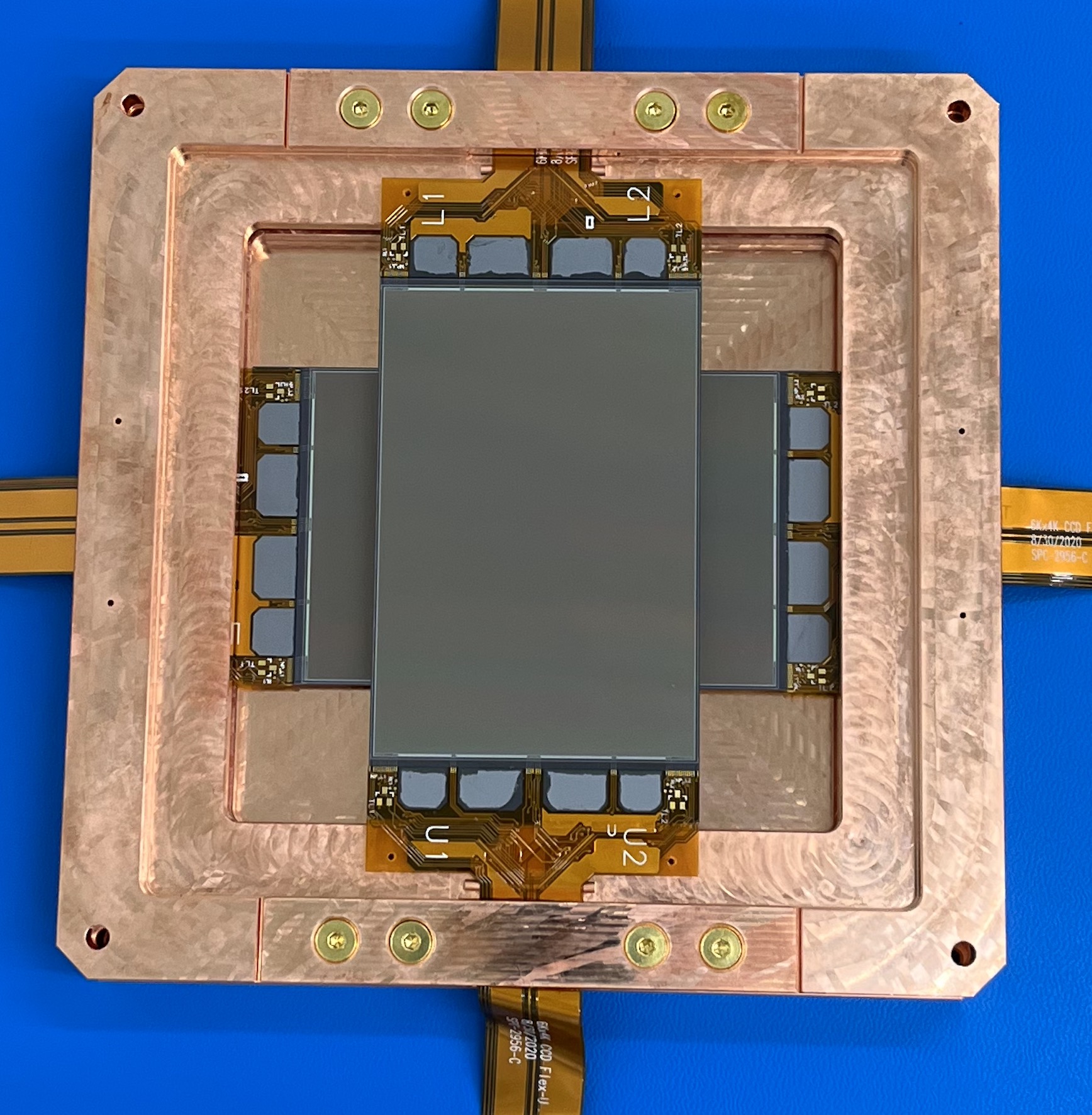}\hfill
    \includegraphics[width=0.5\textwidth, valign=m]{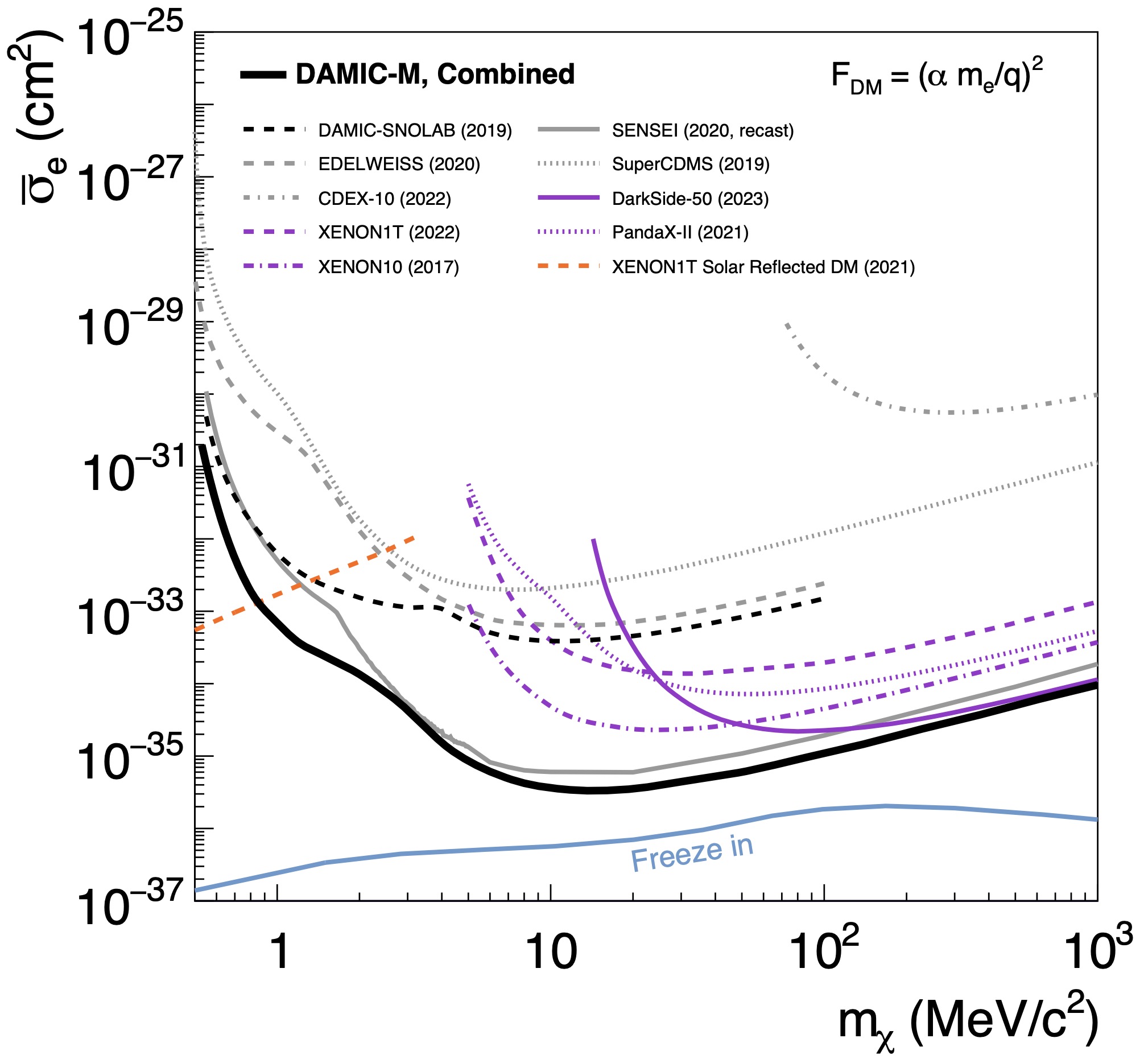}
    \caption{(Left) Two 6k$\times$4k CCDs installed originally in the LBC and used to collect the data for DM-electron studies\,\cite{DAMIC-M:2024ooa}. (Right) DAMIC-M 90\% C.L. upper limits (solid thick black) on DM-electron interactions through an ultralight mediator with limits from other experiments and theoretical expectation from the freeze-in mechanism\,\cite{DAMIC-M:2023hgj}.
    }
    \label{fig:CCDs}
\end{figure*}

\section{Conclusion}

The DAMIC-M project will use precisely calibrated skipper CCDs to push the search for DM into new, unexplored regions that were previously non-accessible due to detector limitations. The DAMIC-M prototype detector, the LBC, has reached the background level achieved in the previous experiment, and its first science data improved existing limits on the DM-electron scattering. After its numerous upgrades, the LBC continues to take science data.

Stringent mitigation steps are being adopted for the DAMIC-M detector to push the background level below 1\,dru in low energies. These days, CCDs are being packaged and tested, electro-formed copper and radio-pure flexes are being produced, and new low-noise electronics is used to run skipper CCDs. The goal is to have the DAMIC-M detector online in early 2025 and search for light DM.


\end{document}